
\input phyzzx
\overfullrule=0pt

\def\rarrow{\rightarrow}
\def\({[}
\def\){]}
\def\e{\epsilon}
\def\t{\eta}

\def\NP{{\it Nucl. Phys.\ }}
\def\PL{{\it Phys. Lett.\ }}
\def\PRL{{\it Phys. Rev. Lett.\ }}
\def\PR{{\it Phys. Rev.\ }}

\REF{\frni}{C.D. Froggatt and H.B. Nielsen, \NP {\bf B147} (1979) 277.}
\REF{\dgg}{A. De Rujula, H. Georgi and S.L. Glashow,
 {\it Ann. Phys.} {\bf 109} (1977) 258.}
\REF{\frit}{H. Fritzsch, \PL {\bf 70B} (1977) 437.}
\REF{\dija}{S. Dimopoulos and C. Jarlskog, \PL {\bf 86B} (1979) 297.}
\REF{\dimo}{S. Dimopoulos, \PL {\bf 129B} (1983) 417;
J. Bagger, S. Dimopoulos, E. Masso and M. Reno,
 \NP {\bf B258} (1985) 565;
J. Bagger, S. Dimopoulos, H. Georgi and S. Raby, In: {\it Proc. Fifth
Workshop on Grand Unification.} Eds. Kang, K., Fried, H. and Frampton,
P., Singapore, World Scientific (1984).}
\REF{\caha}{R. Cahn and H. Harari, \NP {\bf B176} (1980) 135.}
\REF{\gkmr}{B.R. Greene, K.H. Kirklin, P.J. Miron and G.G. Ross,
 \NP {\bf B292} (1987) 606.}
\REF{\drw}{A. Davidson, S. Ranfone and K.C. Wali,
 \PR {\bf D41} (1990) 208.}
\REF{\dhr}{S. Dimopoulos, L.J. Hall and S. Raby,
 \PRL {\bf 68} (1992) 1984; \PR {\bf D45} (1992) 4192.}
\REF{\bamo}{B.S. Balakrishna and R.N. Mohapatra,
 \PL {\bf B216} (1989) 349.}
\REF{\kapl}{ D. Kaplan, \NP {\bf B365} (1991) 259.}
\REF{\morel}{Z.G. Berezhiani and R. Rattazi, LBL-32889 (1992).}
\REF{\ahr}{A. Antaramian, L.J. Hall and A. Rasin,
 \PRL {\bf 69} (1992) 1871.}
\REF{\glwe}{S.L. Glashow and S. Weinberg, \PR {\bf D15} (1977) 1958.}
\REF{\gms}{R. Gatto, G. Morchio and F. Strocchi,
 \PL {\bf 83B} (1979) 348.}
\REF{\gmss}{R. Gatto, G. Morchio, G. Sartori and F. Strocchi,
 \NP {\bf B163} (1980) 221.}
\REF{\sewe}{G. Segre and H.A. Weldon, \PL {\bf 86B} (1979) 291;
 {\it Ann. Phys.} {\bf 124} (1980) 37.}
\REF{\waca}{M.Y. Wang and E.D. Carlson, HUTP-92/A062 (1992).}
\REF{\ghkd}{J.F. Gunion, H.E. Haber, G. Kane and S. Dawson, {\it
The Higgs Hunter's Guide}, (Addison-Wesley, 1990).}
\REF{\hash}{H.E. Haber and M. Sher, \PR {\bf D35} (1987) 2206.}
\REF{\kkw}{G.L. Kane, C. Kolda and J.D. Wells, UM-TH-92-24 (1992).}
\REF{\nisa}{Y. Nir and U. Sarid, WIS-92/52/Jun-PH (1992), \PR {\bf D},
 in press.}
\REF{\gale}{J. Gasser and H. Leutwyler,
 {\it Phys. Rep.} {\bf 87} (1982) 77.}
\REF{\kama}{D. Kaplan and A. Manohar, \PRL {\bf 56} (1986) 2004.}
\REF{\dowy}{J. Donoghue and D. Wyler, \PR {\bf D45} (1992) 892.}
\REF{\choi}{K. Choi, \NP {\bf B383} (1992) 58.}
\REF{\leut}{H. Leutwyler, \NP {\bf B337} (1990) 108.}
\REF{\tom}{T. Banks, \NP {\bf B303} (1988) 172.}
\REF{\wili}{B. McWilliams and L-F. Li, \NP {\bf B179} (1981) 62.}
\REF{\shan}{O. Shanker, \NP {\bf B206} (1982) 253.}
\REF{\cheng}{T.P. Cheng and M. Sher, \PR {\bf D35} (1987) 3484.}
\REF{\hani}{H.E. Haber and Y. Nir, \NP {\bf B335} (1990) 363.}
\REF{\nrt}{E. Nardi, E. Roulet and D. Tommasini,
 \PR {\bf D46} (1992) 3040.}

\def\Weizmann{\centerline{\it Physics Department}
  \centerline{\it Weizmann Institute of Science}
  \centerline{ Rehovot 76100, Israel}}
\def\Rutgers{\centerline{\it Department of Physics and Astronomy}
\centerline{\it Rutgers University}
\centerline{\it Piscataway, NJ 08855-0849, USA}}
{\baselineskip=11pt
\Pubnum={RU-92-59 \cr WIS-92/94/Dec-PH}
\date={December, 1992}
\titlepage
\title{{\bf Mass Matrix Models}}
\author{Miriam Leurer, Yosef Nir}
\Weizmann
\andauthor{Nathan Seiberg}
\Rutgers

\vskip 1.0in

\centerline{\bf Abstract}
It is possible that the hierarchy in the masses and mixing of
quarks is a result of a horizontal symmetry. The smallness of
various parameters is related to their suppression by high powers
of a scale of new physics. We analyze in detail the structure of
such symmetries in view of new experimental data and present
explicit models consistent with all phenomenological constraints.
We show that it is possible that the flavor dynamics can be at
accessible energies -- as low as a $TeV$.
\endpage
}

\chapter{Introduction}

The experimental values of the quark sector parameters, even if
consistent with each other, may provide important hints for new physics
beyond the standard model. In recent years, much progress has been
achieved in the determination of these parameters: heavy quark symmetry
and additional experimental measurements allowed a more accurate
determination of $|V_{cb}|=0.040\pm0.007$; measurements in CLEO and
ARGUS of the leptonic spectrum near the endpoint in charmless $B$ decays
provided a first determination of $|V_{ub}/V_{cb}|=0.10\pm0.03$; and the
combination of direct searches in CDF with precision electroweak
measurements in LEP gave stringent bounds on $m_t\approx135\pm45\ GeV$.
Consequently, all six quark masses and three mixing angles became known
to within a factor of two or so. This calls for a re-examination of
various frameworks that try to explain the hierarchy in these
parameters.
\par
Most theoretical frameworks that provide such an explanation suggest
that the new physics responsible for the hierarchy takes place at a very
high energy scale, typically the GUT scale (see, for example, refs.
$\(\frni-\dhr\)$).  If this is indeed the case, it would be very
difficult to test these ideas beyond the numerical estimates that they
provide for the quark masses and mixings. If, on the other hand, the
physics of flavor lies at the $TeV$ scale (see, for example, refs.
$\(\bamo-\morel\)$) then it may be accompanied with rich phenomenology
that may be accessible to future experiments.
\par
In this work, we study the possibility that the hierarchy in the quark
sector parameters is a result of a horizontal discrete symmetry. We
analyze in detail the structure of such symmetries in view of the new
experimental data; and we carefully examine the possibility that the
symmetry is broken at low enough energies to be tested in experiment.

 We have in mind supersymmetric models.  Most of our discussion
is in this context and we ignore the soft breaking terms.  However,
our ideas are more general and apply also to non-supersymmetric models.

In sections 2--4 we discuss general features of horizontal symmetries
and their breaking mechanism: in section 2 we prove that horizontal
symmetries cannot be exact and require that the scalar sector is
extended beyond the single Higgs doublet of the standard model; in
section 3 we discuss the phenomenological consequences of the symmetry
breaking at various scales and analyze the implications of natural
flavor conservation; and in section 4 we show that only Abelian
horizontal symmetries are relevant in a large class of models.  In
sections 5--6 we introduce a more specific framework, where mixing and
hierarchy of masses arise from non-renormalizable terms: in section 5 we
discuss the theoretical framework and in section 6 we present the
requirements that arise from the measured values of the quark sector
parameters. In section 7 we present several explicit models that explain
the observed hierarchy in the quark sector parameters, and in section 8
we study the possibility that the new physics related to these models
takes place at a low enough scale to be observed in experiment. Finally,
we give our conclusions in section 9.  In an appendix we extend our
ideas to the lepton sector.

\chapter{General Features of Horizontal Symmetries}
Our basic assumption in this work is that there is a horizontal symmetry
$H$ that gives a certain structure to the quark mass matrices.
Throughout our discussion, we assume that $H$ is a discrete symmetry
which might or might not be gauged.  We note that spontaneously broken
discrete symmetries may cause cosmological problems by creating domain
walls.  We have nothing to add in this respect.
\par
In this section, we prove two well known facts about horizontal
symmetries: that they cannot be exact and that they require extending
the scalar sector beyond the single Higgs doublet of the standard model.
We present the proofs in a way which is useful for our argumentation
later.
\par
We denote quark fields by $Q$, $\bar d$ and $\bar u$ for, respectively,
$(3,2)_{1/6}$, $(\bar3,1)_{1/3}$ and $(\bar3,1)_{-2/3}$ representations
of $SU(3)_C\times SU(2)_L\times U(1)_Y$.  Scalar fields are generically
denoted by $\phi$.  (Later we will distinguish between $\phi_d$,
$\phi_u$ and $S$ which denote, respectively, $(2)_{-1/2}$, $(2)_{+1/2}$
and $(1)_0$ representations of $SU(2)_L\times U(1)_Y$.) Yukawa terms are
then of the general form
$${\cal L}_{\rm Yukawa}=\lambda_{ij}^{da} Q_i \phi^a \bar d_j+
\lambda_{ij}^{ub} Q_i \phi^b \bar u_j+{\rm h.c.}.\eqn\yukawa$$
The mass matrices $M_u=\lambda^{ub}\VEV{\phi^b}$ and
$M_d=\lambda^{da}\VEV{\phi^a}$ are diagonalized by bi-unitary
transformations,
$$\eqalign{ U_LM_uU_R^\dagger=&D_u\equiv{\rm diag}(m_u,m_c,m_t),\cr
D_LM_dD_R^\dagger=&D_d\equiv{\rm diag}(m_d,m_s,m_b),\cr }\eqn\nfca$$
with the CKM mixing matrix
$$V=U_LD_L^\dagger.\eqn\nfcb$$
Under $H$, the fields transform as
$$Q\rarrow LQ,\ \ \bar d\rarrow R_d\bar d,\ \ \bar u\rarrow R_u\bar u,
\ \ \phi\rarrow P\phi.\eqn\nfcc$$
In the mass basis, left handed quarks transform as
$$d\rarrow L_dd,\ \ u\rarrow L_uu,\eqn\nfclh$$
where
$$L_u=U_LLU_L^\dagger,\ \ \ L_d=D_LLD_L^\dagger.\eqn\nfcd$$
Note that $L_u$ and $L_d$ are related through
$$L_d=V^\dagger L_u V.\eqn\nfcf$$
\par
First, we prove that {\it if $H$ is not broken, there are
either degenerate quarks or vanishing mixing angles.} If $H$
is unbroken, then the mass matrices are invariant under the symmetry
operation, namely
$$L M_dR_d=M_d,\ \ \ L M_uR_u=M_u,\eqn\nfcea$$
and consequently
$$L M_dM_d^\dagger L^\dagger=M_dM_d^\dagger,\ \ \ L
M_uM_u^\dagger L^\dagger=M_uM_u^\dagger.\eqn\nfce$$
In the mass basis, the conditions \nfce\ translate into
$$\(L_d,D_d^2\)=\(L_u,D_u^2\)=0.\eqn\nfcfa$$
Then the fact that there is
no degeneracy in either quark sector forces $L_d$ and $L_u$ to be
diagonal. However, eq. \nfcf\ implies that $L_d$ and $L_u$ have the same
eigenvalues. If the three eigenvalues are different {}from each other,
then \nfcf\ requires that all mixing angles vanish ($V$ is a permutation
$\times$ phase matrix). If two eigenvalues are equal but different
{}from the third one, then one mixing angle vanishes ($V$ is block
diagonal). The fact that none of the mixing angles vanishes forces
$L_d$ and $L_u$ to be proportional to the unit matrix and $H$ is just
an overall phase transformation.

More intuitively, if the horizontal symmetry is not broken, then quarks
in the same representation remain degenerate.  Since there is no
degeneracy in the spectrum, the quarks should be in three one
dimensional representations.  If these representations are different,
there cannot be mixing between them and if they are all isomorphic
representations, the horizontal symmetry is trivial.  We conclude: as
there is no degeneracy among quarks and as all three quark generations
mix, there can be no unbroken horizontal symmetry.
\par
Second, we prove that {\it if $H$ is broken by the
VEV of a single Higgs doublet, there are either degenerate quarks or
vanishing mixing angles.} (The situation in supersymmetric theories with
two Higgs fields will be discussed in the next section.)
If there is a single Higgs doublet, it is in a one dimensional
representation of $H$
which is just a phase transformation.  Therefore, we can always find a
symmetry $H^\prime\subset H\times U(1)_Y$ isomorphic to $H$
under which the Higgs field is neutral.
Note that $H^\prime$ is an exact horizontal symmetry of the full theory
and is {\it not} spontaneously broken. Then, as in the case of unbroken
$H$, it implies either degenerate quarks or trivial mixing angles.
\par
We conclude: as there is no degeneracy among quarks and as all three
quark generations mix, a horizontal symmetry requires extending the
scalar sector beyond a single Higgs doublet.

\chapter{Spontaneous Breaking of Horizontal Symmetries}
{}From the phenomenological point of view, the scale $\Lambda_H$ at
which the spontaneous breakdown of the horizontal symmetry takes place
can reside in one of three ranges:
\par
\noindent
$(i)$ High $\Lambda_H$: above $10^3-10^4\ TeV$. In this range there are
no obvious phenomenological signatures to the existence of $H$, except,
of course, for the numerical estimates of the quark sector parameters
that it provides.
\par
\noindent
$(ii)$ Intermediate $\Lambda_H$: above a few $TeV$ and below
$10^3-10^4\ TeV$. In this range, the new physics related to $H$ may
affect rare processes such as $K-\bar K$ mixing.
\par
\noindent
$(iii)$ Low $\Lambda_H$: a few $TeV$ or less. In this range,
the new physics related to $H$ may be {\it directly} accessible in
future experiments.
\par
Models with a high $\Lambda_H$ are phenomenologically ``safe": scalar
doublets and singlets have their masses at the scale $\Lambda_H$ except
for one Higgs (two in the SUSY framework).  The light scalar sector is
similar to the minimal (SUSY) standard model.  But it is somewhat
disappointing that all ``flavor physics" takes place at a high energy
scale and cannot be further tested in experiments.
\par
If, on the other hand, $H$ is broken at an intermediate or maybe even at
the electroweak scale, then the implications for phenomenology of the
scalar sector are very interesting.  (For a recent related
discussion, see ref. $\(\ahr\)$.)  As we showed in the previous section,
the scalar sector should be extended beyond a single Higgs doublet.
However, a generic multi Higgs model leads to large flavor changing
neutral currents (FCNC).  For nondiagonal Yukawa couplings of order
one, the bounds from neutral meson mixing,
$$M(\phi_i)\gsim\cases{
4\times10^3\ TeV\ \sqrt{\lambda^i_{ds}\lambda^i_{sd}}
&($K-\bar K$ mixing),\cr
6\times10^2\ TeV\ \sqrt{\lambda^i_{uc}\lambda^i_{cu}}
&($D-\bar D$ mixing),\cr
6\times10^2\ TeV\ \sqrt{\lambda^i_{db}\lambda^i_{bd}}
&($B-\bar B$ mixing),\cr
}\eqn\nfci$$
do not allow intermediate or low $\Lambda_H$.
\par
A well known mechanism to avoid FCNC is that of Natural Flavor
Conservation (NFC) $\(\glwe\)$, which requires that there is only one
scalar doublet coupled to each quark sector:
$${\cal L}_{\rm Yukawa}=\lambda_{ij}^d Q_i\phi_d \bar d_j
+\lambda_{ij}^u Q_i\phi_u \bar u_j+{\rm h.c.}.\eqn\nfcyuk$$
This is the situation in supersymmetric models.
An important characteristic feature of
these couplings is the existence of a global $U(1)_X$ symmetry:
$$\bar d\rarrow e^{i\alpha}\bar d,\ \ \phi_d\rarrow e^{-i\alpha}\phi_d,
\eqn\xsym$$
and all other light fields carry $X=0$.  Note that $U(1)_X$ is not
necessarily a symmetry of the full Lagrangian but only of the terms in
equation \nfcyuk. Such a situation with an extended symmetry of the
Yukawa terms is natural either in supersymmetric models or with suitably
chosen discrete symmetries.
The Yukawa Lagrangian \nfcyuk\ is, however,
unacceptable in the framework of horizontal symmetry: The combination of
spontaneously broken horizontal symmetry and NFC leads to either
degenerate quarks or trivial mixing angles $\(\gms-\sewe\)$.
\par
To prove this statement, note that as there is only one scalar doublet
coupled to each quark sector, each of them is in a one dimensional
representation of $H$. The transformation law \nfcc\ can be rewritten
for the scalar fields:
$$\phi_d\rarrow P_d\phi_d,\ \ \ \phi_u\rarrow P_u\phi_u,\eqn\nfcg$$
where $P_q$ is a phase transformation.  Then we
can define a symmetry $H^\prime$ of the Yukawa terms \nfcyuk, isomorphic
to $H$, that operates on the various fields as
$$\eqalign{Q\rarrow&LQ,\ \ \bar d\rarrow R_dP_d\bar d,\ \
\bar u\rarrow R_uP_u\bar u,\cr
\phi_d\rarrow&\phi_d,\ \ \phi_u\rarrow \phi_u.\cr}\eqn\nfch$$
$H^\prime$ is a subgroup of $H\times U(1)_X\times U(1)_Y\times U(1)_B$
($U(1)_B$ is baryon number symmetry).  Such redefinitions of the
horizontal symmetry will be used extensively below.  Since both
$\phi_u$ and $\phi_d$ transform trivially under $H^\prime$, $H^\prime$
is not broken. As proved in the previous section, this implies either
degenerate quarks or trivial mixing angles.
\par
 This conclusion is not modified even if there are additional scalars
that transform nontrivially under $H$ but do not couple to quarks: we
define all of them to be singlets of $H^\prime$.  Note that even though
$U(1)_X$ is not necessarily a symmetry of the full Lagrangian,
$H^\prime$ is.
\par
Thus, to allow for an intermediate or low $\Lambda_H$, we would like to
find a framework where NFC is broken, but by a small amount.  The no-go
proof above points at a way around it.  One could violate NFC by
non-renormalizable terms.  They are naturally small because they are
suppressed by inverse powers of some high scale $M$. If there is
only one Higgs, the result in the previous section shows that it is
impossible to break the horizontal symmetry.  With only two Higgs
fields, $\phi_u$ and $\phi_d$, the horizontal symmetry cannot be broken,
even when non-renormalizable operators are taken into account, without
violating the $U(1)_X$ symmetry.  We are, therefore, led to consider
terms like $Q\phi_d (\phi_u\phi_d)^n \bar d$ and $Q\phi_u
(\phi_u\phi_d)^n \bar u$ which violate this symmetry and lead to flavor
changing processes.  Alternatively, we can preserve the $U(1)_X$
symmetry in the Yukawa couplings and the non-renormalizable terms
but introduce more scalars, $S$, which
are invariant under $SU(2)\times U(1)$ with $U(1)_X$ invariant
couplings like $Q\phi_d S^n \bar d$ and $Q\phi_u S^n \bar u$.
\par
The framework of this work is then defined as follows: We assume that
there is a horizontal symmetry $H$ which is spontaneously broken,
preferably at a low scale $\Lambda_H$. In addition, there is Natural
Flavor Conservation broken by non-renormalizable terms and suppressed,
therefore, by a high energy scale $M\ >\ \Lambda_H$.
\chapter{Abelian or nonabelian?}
Within our framework, the tree level renormalizable Yukawa terms are of
the form \nfcyuk. Then $\phi_d$ and $\phi_u$ are necessarily in one
dimensional representations of $H$.
{\it If all the scalars that couple
to quarks are in one dimensional representations of $H$, then all quarks
are also in one dimensional representations}\foot{A
similar point was recently made in a different context in ref.
$\(\waca\)$.}.
\par
To prove this statement, note that as the scalars have only phase
transformations under $H$, we must have
$$\(L,\lambda^u\lambda^{u\dagger}\)=\(L,\lambda^d\lambda^{d\dagger}\)
=0.\eqn\nfcj$$
For three quark generations, there are three possibilities:
\par
\noindent
($i$) $L$ is an irreducible representation of $H$.  Then eq. \nfcj\
implies that both $\lambda^u\lambda^{u\dagger}$ and
$\lambda^d\lambda^{d\dagger}$ are proportional to the unit matrix,
leading to degeneracy among all three quarks. Such a degeneracy cannot
be lifted enough by small corrections.
\par
\noindent
($ii$) $L$ is reducible to a two dimensional and a one dimensional
representations. Then \nfcj\ implies that the two generations in the two
dimensional representation are degenerate. But more important, the
single generation in one dimensional representation does not mix with
the other two generations. This holds to all orders since any
combination of scalars is still in one dimensional representation of
$H$.

A way out of this scenario is to add an $SU(2)$-singlet scalar in a two
dimensional representation of $H$. Then nonrenormalizable terms may
induce the required mixing.
\par
\noindent
($iii$) $L$ consists of three one dimensional representations.
This implies that $R_d$ and $R_u$ also reduce to one dimensional
representations.
\par
The conclusion is that if all scalars reside in one dimensional
representations of a horizontal symmetry, then only an Abelian
(sub-)group $H$ is relevant.

\chapter{Mass and Mixing Hierarchy from Nonrenormalizable Terms}
The mechanism that we employ to create the hierarchy in the masses and
mixing of quarks is similar to that suggested by Froggatt and Nielsen
$\(\frni\)$.  To explain the general idea, it is simplest to consider
the spontaneous breaking of the horizontal symmetry by a VEV of a scalar
$S$ which is a singlet of the standard model. Then the breaking scale is
simply
$$\Lambda_H\sim\VEV{S}.\eqn\nta$$
We further assume that at some higher energy scale, $M$, natural flavor
conservation is broken. Specifically, we will introduce additional pairs
of mirror fermions with masses of order $M$.
\par
Consider first the possibility that the $U(1)_X$ symmetry of \xsym\ can
act also on the massive fermions such that all Yukawa terms in the
Lagrangian are invariant.  When we integrate out the massive fermions
the effective Yukawa couplings below $M$ are of the form
$$ {\lambda_{ij}^d\over
M^{m_{ij}}}Q_i\phi_d S^{m_{ij}}\bar d_j+ {\lambda_{ij}^u\over
M^{n_{ij}}}Q_i\phi_u S^{n_{ij}}\bar u_j +{\rm h.c.}.\eqn\ntb$$
In non-supersymmetric models we should also allow powers of
$S^\dagger$.  They can be incorporated in the following discussion by
allowing negative $m$ and $n$ but we will ignore this possibility.
Without loss of generality, we can use $U(1)_Y\times U(1)_X$ rotations
and rescaling to redefine the horizontal charges (which we also denote
by $H$) of the scalar fields to
$$H(\phi_d)=H(\phi_d)=0,\ \ \ H(S)=-1.\eqn\ntc$$
Then the power $m_{ij}(n_{ij})$ in eq. \ntb\ is simply the horizontal
charge difference between $Q_i$ and $d_j(u_j)$:
$$m_{ij}=H(Q_i)+H(\bar
d_j),\ \ \ n_{ij}=H(Q_i)+H(\bar u_j).\eqn\ntd$$
(We assume for
simplicity that the horizontal symmetry is $U(1)_H$ rather than a
discrete subgroup of it.  Suitable terms in the scalar potential
explicitly break $U(1)_H$ to a discrete subgroup.  As mentioned above,
such a situation can be natural.)  In the quark mass matrix,
$${\cal L}_{\rm mass}=\lambda_{ij}^d\VEV{\phi_d}\left( {\Lambda_H\over
M}\right)^{m_{ij}}Q_i\bar d_j+\lambda_{ij}^u\VEV{\phi_u}\left({\Lambda_H
\over M}\right)^{n_{ij}}Q_i\bar u_j+{\rm h.c.},\eqn\ntda$$
a small parameter
$$\epsilon={\Lambda_H\over M}.\eqn\nte$$
appears (we absorb a Yukawa coupling of $S$ to massive
fields into the definition of $\Lambda_H$).  {\it The hierarchy in the
quark sector parameters appears because various mixing angles and mass
ratios depend on different powers $\epsilon^m$.} Even if $\epsilon$ by
itself is not a very small number, the physical parameters may be very
small if they depend on high powers $m$, namely if there are large
$H$-charge differences between quarks.
\par
The mixing angles are determined by the $H$-charge differences between
the quark doublets:
$$|V_{us}|\sim\e^{\(H(Q_1)-H(Q_2)\)},\ \
|V_{cb}|\sim\e^{\(H(Q_2)-H(Q_3)\)},\ \
|V_{ub}|\sim\e^{\(H(Q_1)-H(Q_3)\)}.\eqn\frnia$$
Eq. \frnia\ implies that
$$|V_{ub}|\sim|V_{us}V_{cb}|\eqn\frnib$$
always holds in our framework.
As for the mass ratios, they depend on the charges [\frni]:
$${m_{d_i}\over m_{d_j}}\sim
\e^{\(H(Q_i)-H(Q_j)+H(\bar d_i)-H(\bar d_j)\)},\
\ {m_{u_i}\over m_{u_j}}\sim
\e^{\(H(Q_i)-H(Q_j)+H(\bar u_i)-H(\bar u_j)\)}.\eqn\frnic$$
\par
Note that with the Yukawa Lagrangian \ntb, there are no flavor changing
neutral currents mediated by the scalar doublet fields. The effective
Yukawa couplings of $\phi_d$,
$(\lambda^d_{eff})_{ij}=\lambda^d_{ij}({\VEV{S}\over M})^{m_{ij}}$, and
of $\phi_u$, $(\lambda^u_{eff})_{ij}=\lambda^u_{ij}({\VEV{S}\over
M})^{n_{ij}}$, can be diagonalized simultaneously with the
corresponding mass matrices.  Only the singlet scalar $S$ mediates FCNC.
This is rather fortunate: lower bounds on the doublet masses {}from FCNC
may be difficult to satisfy because they may be in conflict with upper
bounds {}from triviality or unitarity. In particular, in a
supersymmetric framework, there is in general a strong upper bound on
the mass of the lightest neutral scalar\foot{We thank T. Banks for
stressing the relevance of these bounds to our problem.}, of order $150\
GeV$ (see {\it e.g.} refs.  $\(\ghkd-\kkw\)$). On the other hand, the
mass of a singlet scalar is not related to the electroweak breaking
scale and could be easily set to satisfy constraints from FCNC.
We will study these constraints in section 8.

Next, we consider the possibility that $U(1)_X$ of eq. \xsym\
is not a symmetry of the full Lagrangian and this leads to
Yukawa interactions of the form
$$ {\lambda_{ij}^d\over M^{m_{ij}+2k_{ij}}}Q_i\phi_d(\phi_d\phi_u)^
{k_{ij}}S^{m_{ij}}\bar d_j+{\lambda_{ij}^u\over M^{n_{ij}+2l_{ij}}}
Q_i\phi_u(\phi_d\phi_d)^{l_{ij}}S^{n_{ij}}\bar u_j+{\rm h.c.}
\eqn\ntf$$
with non-zero $k_{ij}$ and $l_{ij}$.
This has two important implications:
\item{a.}
There is an additional small parameter,
$$\eta^2={\VEV{\phi_d}\VEV{\phi_u}\over M^2}.\eqn\ntg$$
Together with $\epsilon$, a more detailed explanation of the hierarchy
in parameters may become possible.
\item{b.}
There are FCNC mediated by the doublet scalar fields (and not only by
the singlet). This may lead, as discussed above, to phenomenological
problems in some cases.
\par
{}From \ntf\ it can be seen
that we can do with no singlets at all, with just $\eta^2$ as our small
parameter. This scenario, however, would turn out to lead to
considerable difficulties.

Three comments are in order.
\par
\noindent
$(i)$   Another generalization (which we will study in detail below) is
based on adding more singlets.
For example, with two singlets, $S_1$ and $S_2$,
there are two small parameters (even when $U(1)_X$ holds at low energy),
$\e_1$ and $\e_2$. This, again, would allow more structure to the mass
matrices though some of the predictivity is lost.

\noindent
$(ii)$  In general $U(1)_H\times U(1)_X$
is explicitly broken by the scalar potential and the horizontal symmetry
of the full Lagrangian is a discrete subgroup
${\cal H} \subset U(1)_H \times
U(1)_X$. It is easy to construct models such that $\cal H$ invariance
guarantees that the Yukawa couplings are invariant under $U(1)_H \times
U(1)_X$.

\noindent
$(iii)$ The expressions for the mixing angles \frnia\ and for the
masses \frnic\ imply that, if $H(\bar d_i)-H(\bar d_j)=H(Q_i)-H(Q_j)$,
then the corresponding $2\times2$ mass matrix has the form
$$M\sim\pmatrix{\e^{2m} & \e^m \cr \e^m & 1}\ \Longrightarrow\
|V_{ij}|\sim\sqrt{m_i\over m_j}.\eqn\nth$$
It is well known that such a relation holds to within a few
percent for the two light generations, $|V_{us}|\approx\sqrt{m_d\over
m_s}$. We emphasize that while the order of magnitude relation
\nth\ is easy to derive in our framework, an actual equality
as the experimental values may suggest (see next section) is
difficult to obtain. The reason is that the natural way to
obtain an equality $\(\frit\)$,
$$M=\pmatrix{0 & a\e^m \cr a\e^m & b}\ \Longrightarrow\
|V_{ij}|=\sqrt{m_i\over m_j},\eqn\nti$$
requires not only $M_{11}=0$ but also $|M_{12}|=|M_{21}|$. The latter
equality is not natural without left-right symmetry, but we were
unable to find left-right symmetric models consistent with all
our requirements.

\chapter{Numerology}
In this section we introduce the experimental values of the quark sector
parameters and analyze the hierarchy required in the mass matrices to
produce these values.
\par
For the mixing angles, we take (see ref. $\(\nisa\)$ and references
therein)
$$|V_{us}|=0.2205\pm0.0018,\ \ |V_{cb}|=0.040\pm0.007,\ \
|V_{ub}/V_{cb}|=0.10\pm0.03.\eqn\mix$$
(The complex KM phase in the mixing matrix is not a small parameter,
$\sin\delta={\cal O}(1).$ We do not discuss CP violation in this work.)
Note that the relation \frnib, $|V_{ub}/V_{cb}|\sim|V_{us}|$ holds to
within a factor of 2--3. This is encouraging, because the relation is a
very general feature of our framework.
\par
For quark masses {\it at 1 GeV}, we take $\(\gale\)$
$$\eqalign{m_u&=5.1\pm1.5\ MeV,\ \
m_c=1.35\pm0.05\ GeV,\ \ m_t\sim225\pm75\ GeV,\cr
m_d&=8.9\pm2.6\ MeV,\ \
m_s=175\pm55\ MeV,\ \ m_b=5.6\pm0.4\ GeV,
\cr}\eqn\mass$$
(we used $m_t^{\rm phys}\approx0.6m_t(1\ GeV)$) leading to the mass
ratios:
$$\eqalign{
{m_d\over m_s}=&0.051\pm0.004,\ \ \
{m_s\over m_b}=0.032\pm0.012,\cr
{m_u\over m_c}=&0.0038\pm0.0012,\ \
{m_c\over m_t}\sim0.006^{+0.003}_{-0.002},\cr
{m_b\over m_t}\sim&0.025^{+0.015}_{-0.008}.\cr
}\eqn\ratios$$
\par The largest of the small parameters is the Cabibbo angle,
$|V_{us}|\sim0.22$.  If we set our small expansion parameter to equal
$|V_{us}|$, then we should roughly aim at
$$\eqalign{\e&\sim|V_{us}|,\cr
\e^2&\sim|V_{cb}|,\ {m_d\over m_s},\ {m_s\over m_b},
\ {m_b\over m_t},\cr
\e^3&\sim|V_{ub}|,\ {m_u\over m_c},\ {m_c\over m_t}.\cr}\eqn\ideal$$
With this order of magnitude estimate, we have
$$\det M_d\sim \VEV{\phi_d}^3 \e^{12},\ \ \ \det M_u\sim \VEV{\phi_u}^3
\e^9 \eqn\detmq$$
where $\VEV{\phi_d}$ is of the same order as $\VEV{\phi_u}$.  As we will
see below, the very high powers of $\e$ mean that we would need a large
number of fields in the high energy theory to produce this hierarchy.
\par
If we allow $m_u=0$ (a possibility which is still controversial
$\(\kama-\leut\)$), the determinant of the block of massive $u$ quarks
is $m_t m_c\sim \VEV{\phi_u}^2 \e^3$.  If we explain the small ratio
$m_b/m_t$ dynamically, namely $m_b/m_t\sim\VEV{\phi_d}/\VEV{\phi_u}$ as
in [\tom],
rather than obtaining the hierarchy from the Yukawa couplings, then
$\det M_d\sim \VEV{\phi_d}^3\e^6$.  In either of these possibilities the
required high energy model could have fewer massive fields.
\par
As we will later see, it is difficult to construct a viable low energy
model with $\epsilon$ as large as 0.22. We may need two small
parameters, $\e_1 \sim 0.04$ and $\e_2\sim 0.008$:
$$\eqalign{
\e_1&\sim|V_{cb}|,\ {m_s\over m_b},
\ {m_b\over m_t},\cr
\e_2&\sim|V_{ub}|,\ {m_u\over m_c},\ {m_c\over m_t},\cr
{\e_2\over\e_1}&\sim|V_{us}|,\ \sqrt{m_d\over m_s}.\cr
}\eqn\idealb$$
With this order of magnitude estimate, we have
$$\det M_d\sim \VEV{\phi_d}^3 \e_1^3\e_2^2,\ \ \ \det M_u\sim
\VEV{\phi_u}^3 \e_2^3.\eqn\detmqb$$
If $m_b/m_t\sim\VEV{\phi_d}/\VEV{\phi_u}$ then $\det M_d\sim
\VEV{\phi_d}^3\e_2^2$.
If $m_u=0$ then $m_tm_c\sim \VEV{\phi_u}^2\e_2$.
\par
Finally, we mention the possibility that there is only one small
parameter, $\e \sim 0.04$ and a much cruder estimate of the various
parameters can be achieved:
$$\eqalign{|V_{us}|&\sim1\ {\rm or}\ \e,\cr
|V_{cb}|&,\ {m_d\over m_s},\ {m_s\over m_b},
\ {m_b\over m_t}\sim\e,\cr
|V_{ub}|&,\ {m_u\over m_c},\ {m_c\over m_t}\sim\e\ {\rm or}\ \e^2 .\cr}
\eqn\idealc$$

\chapter{Explicit Examples}
We now turn to some explicit examples of the general ideas discussed
above.  We first construct a low energy model with two small parameters
$\e_1$ and $\e_2$. We assume that they arise {}from the VEV's of two
singlet fields $S_1$ and $S_2$:
$$\e_1={\VEV{S_1}\over M}\sim0.04,\ \ \
\e_2={\VEV{S_2}\over M}\sim0.008.\eqn\expla$$
These small numbers are not extremely small and are perfectly natural.
As we will see, we absorb a typical Yukawa coupling into the definition
of $\VEV{S_i}$.  This makes $\e$ smaller than the ratio of scales.
Furthermore, the ratio between the two VEV's,
$${\e_2\over\e_1}\sim0.2,\eqn\explb$$
is a number of order one. It can arise dynamically by
minimizing the scalar potential.  We do not address this
issue here.

Let us consider the following set of matrices, where the various entries
give the corresponding orders of magnitude (and not exact numbers):
$$M_u=\VEV{\phi_u}\pmatrix{\e_2^2&0&\e_2\cr
\e_1\e_2&\e_2&\e_1\cr \e_2&0 &1\cr},
\qquad\qquad
M_d=\VEV{\phi_d}\pmatrix{\e_2^2&\e_1\e_2&\e_1\e_2\cr
\e_1\e_2&\e_1^2&\e_1^2\cr \e_2&\e_1&\e_1\cr}.\eqn\explc$$
For $\VEV{\phi_u}\sim\VEV{\phi_d}$ they lead to exactly the required
relations of eq. \idealb.

Such mass matrices can arise from non-renormalizable terms in the low
energy Lagrangian with two fermions and a number of scalars.  We take
the horizontal symmetry to be $U(1)_{H_1}\times U(1)_{H_2} $ and also
impose the $U(1)_X$ symmetry on the Yukawa couplings.
We actually have in mind an anomaly free discrete subgroup of
$U(1)_X\times U(1)_{H_1}\times U(1)_{H_2}$.  For example, we can
consider the symmetry $Z_3\times Z_3
\times Z_{5} \subset U(1)_X\times U(1)_{H_1}\times U(1)_{H_2}$.
Then the models presented below are free of QCD anomalies.  Anomalies
with the $SU(2)$ gauge group cannot be discussed without assigning
charges to the leptons and without deciding whether the horizontal
symmetry is an R-symmetry or not.
All the results concerning the quark mass matrices are unchanged,
so we use the simpler presentation with $U(1)$ charges.

Using $U(1)_Y\times U(1)_X$ we can set the horizontal charges of the
scalars to be
$$\eqalign{
H_1(\phi_u)&=H_1(\phi_d)=H_1(S_2)=0,\ \ H_1(S_1)=-1,\cr
H_2(\phi_u)&=H_2(\phi_d)=H_2(S_1)=0,\ \ H_2(S_2)=-1.\cr}\eqn\explca$$
Using the baryon number symmetry $U(1)_B$ we can set $H_1(Q_3) =
H_2(Q_3)$ $ = 0$.  Then the mass matrices \explc\ are consistent with
the $H_1$ and $H_2$ charges for quarks:
$$\vbox{\settabs\+ {\rm field} $\quad$ &
$\quad- {2 \over 3}$  & $\quad x -2$ \cr
\+ \rm {field} & $ H_1\quad$  & $H_2\quad $ \cr
\+ $Q_1$ & $0$  & $1$ \cr
\+ $Q_2$ & $1$  & $0$ \cr
\+ $Q_3$ & $0$  & $0$ \cr
\+ $\bar u_1$  & $0$ & $1$ \cr
\+ $\bar u_2$  & $-1$ & $1$ \cr
\+ $\bar u_3$  & $0$ & $0$ \cr
\+ $\bar d_1$  & $0$ & $1$ \cr
\+ $\bar d_2$  & $1$ & $0$ \cr
\+ $\bar d_3$  & $1$ & $0$ \cr
}\eqn\expld$$

We can replace the two fields $S_1$ and $S_2$ with a single scalar $S$
such that $\e={\VEV{S}\over M}\sim0.2$ satisfies $\e^3 = \e_2$ and $\e^2
=\e_1$ (thus explaining $\e_2^2=\e_1^3$).  The $U(1)_{H_1}\times
U(1)_{H_2}$ symmetry is replaced then by a single $U(1)_H$ generated by
$H=2H_1+3H_2$, with $H(S)=-1$. This allows also non-zero couplings at
the $(1,2)$ and $(3,2)$ entries of $M_u$.  Clearly, in such a scheme
there is no need to explain the ratio ${\e_2\over\e_1} =0.2$.

We now turn to the high energy theory.  As in the mechanism of [\frni],
we would like to introduce
massive fields such that when they are integrated out we derive the
non-renormalizable terms needed in the low energy theory.  There are
many ways to do that.  For example we can add $SU(2)$ singlet color
triplet fields $U_i$ with charge +2/3 and $D_i$ with charge --1/3:
$$\vbox{\settabs\+ {\rm field} $\quad $ &
$\quad- {2 \over 3}$  & $\quad x -2$ \cr
\+ \rm {field} & $H_1\quad$ & $H_2\quad $ \cr
\+ $U_1$  & $1$ & $0$ \cr
\+ $U_2$  & $0$ & $0$ \cr
\+ $U_3$  & $0$ & $1$ \cr
\+ $D_1$  & $0$ & $0$ \cr
\+ $D_2$  & $0$ & $0$ \cr
\+ $D_3$  & $1$ & $0$ \cr
\+ $D_4$  & $0$ & $1$ \cr
\+ $D_5$  & $-1$& $1$ \cr}\eqn\exple$$
and fields with conjugate horizontal and gauge quantum numbers
$\bar U_i$ and $\bar D_i$.

For the up sector we need a $6\times 6$ matrix whose rows correspond to
$\{Q_1, Q_2, Q_3,$ $ U_1, U_2, U_3\}$ and its columns to $\{ \bar u_1,
\bar u_2, \bar u_3,$ $ \bar U_1, \bar U_2, \bar U_3\}$.  For the down
sector we need an $8 \times 8$ matrix whose rows correspond to $\{Q_1,
Q_2, Q_3, $ $D_1, D_2, D_3, D_4, D_5\}$ and columns to $\{ \bar d_1,
\bar d_2, \bar d_3, $ $\bar D_1, \bar D_2, \bar D_3, \bar D_4, \bar
D_5\}$.  When there are two or more fields with the same quantum
numbers, we use the freedom to mix them to set some of the explicit mass
terms in the matrices below to zero.  Using the quantum numbers we find
the orders of magnitude for the various entries ($M$ is the high scale):
$$M_u^{\rm full}=\pmatrix{0&0&0&0&0&\VEV{\phi_u}\cr
                      0&0&0&\VEV{\phi_u}&0&0\cr
                      0&0&\VEV{\phi_u}&0&\VEV{\phi_u}&0\cr
                      0&\VEV{S_2}&\VEV{S_1}&M&\VEV{S_1}&0\cr
                      \VEV{S_2}&0&0&0&M&0\cr
                      0&0&\VEV{S_2}&0&\VEV{S_2}&M\cr},\eqn\explf$$
$$M_d^{\rm full}=\pmatrix{0&0&0&0&0&0&\VEV{\phi_d}&0\cr
             0&0&0&0&0&\VEV{\phi_d}&0&0\cr
             0&0&0&\VEV{\phi_d}&\VEV{\phi_d}&0&0&0\cr
             \VEV{S_2}&\VEV{S_1}&\VEV{S_1}&M&0 &0&0&0\cr
             \VEV{S_2}&\VEV{S_1}&\VEV{S_1}&0&M &0&0&0\cr
             0&0&0&\VEV{S_1}&\VEV{S_1}&M&0&0\cr
             0&0&0&\VEV{S_2}&\VEV{S_2}&0&M&\VEV{S_1}\cr
             0&\VEV{S_2}&\VEV{S_2}&0&0 &0&0&M\cr}.\eqn\explg$$
(The two fields $D_1$ and $D_2$ with identical quantum numbers are
needed to ensure $\det M_d^{\rm full}\neq0$.)
In finding eqs. \explf\ and \explg\ we used the fact that the theory is
supersymmetric and there are no terms proportional to $S_i^\dagger$.
After integrating out the
massive fields we find for the light three quark generations precisely
the mass matrices \explc.

Although we can have different sets of fields in the high energy theory
leading to an acceptable mass matrix for the light fields, a simple
argument shows that we need at least three $U$ and at least five $D$
quarks (we can also replace a pair of $SU(2)$ singlets $U$ and $D$ with
a single colored doublet).  The mass matrices \explc\ have
$$\eqalign{\det M_u=&
\VEV{\phi_u}^3\e_2^3=\VEV{\phi_u}^3\VEV{S_2}^3 M^{-3},\cr
\det M_d=&\VEV{\phi_d}^3\e_1^3\e_2^2=
\VEV{\phi_d}^3\VEV{S_1}^3\VEV{S_2}^2 M^{-5}.\cr}\eqn\ecplh$$
Since the determinants of $M_u^{\rm full}$ and $M_d^{\rm full}$ are
polynomials in $\VEV{\phi_q}$, $\VEV{S_i}$ and $M$ and they differ
{}from the determinants of $M_u$ and $M_d$ by powers of $M$, we need at
least three $U$ quarks and at least five $D$ quarks.  This argument is
very general.  It is independent of the gauge and horizontal quantum
numbers of the massive fermions.  It also counts correctly different
fermions with identical quantum numbers which are needed to make the
determinant non-zero.

If $m_u=0$, we can set some of the entries in $M_u$ to zero.  The
determinant of the block of massive light particles is then
$\VEV{\phi_u}^2\e_2$.  By the previous argument such a matrix can be
obtained with only one $U$ quark in the high energy theory.  An example
of such a theory is the previously discussed theory but without $U_2$
and $U_3$.

If the smallness of $m_b/m_t$ arises from a small ratio of VEV's
$\VEV{\phi_d}/\VEV{\phi_u}$, we can have $m_b\sim\VEV{\phi_d}$ without a
small $\e$ parameter.  In this case $\det M_d=\VEV{\phi_d}^3\e_2^2$.
Such a matrix can be obtained with only two $D$ quarks in the massive
fermion sector.  An example of such a model is derived by modifying the
$H_1$ and $H_2$ charges of the singlet down quarks of the model above to
$$\vbox{\settabs\+ {\rm field} $\quad$ &
$\quad- {2 \over 3}$  & $\quad x -2$ \cr
\+ \rm {field} & $ H_1\quad$  & $H_2\quad $ \cr
\+ $\bar d_1$  & $-1$ & $1$ \cr
\+ $\bar d_2$  & $0$ & $0$ \cr
\+ $\bar d_3$  & $0$ & $0$ \cr}\eqn\expli$$
and by dropping the massive fermions $D_1$, $D_2$ and $D_5$.  After
integrating out the two heavy fields we find
$$M_d=\VEV{\phi_d}\pmatrix{0&\e_2&\e_2\cr \e_2&\e_1&\e_1\cr
0&1&1\cr}.\eqn\explj$$
\chapter{The Scales in the Problem}
Within our theoretical framework, there are three energy scales:
the electro-weak breaking scale $\sim\VEV{\phi_u}$, the horizontal
symmetry breaking scale $\sim\Lambda_H\sim \VEV{S}$, and the mass scale
of the extra massive fermions $\sim M$. In the explicit models
presented in the previous section, neither $\VEV{S}$ nor
$M$ are fixed, but their ratio $\VEV{S}/M$ is constrained to be
of order 0.04.
It is then possible that the new physics related to
the scales $\VEV{S}$ and $M$ may be at energies accessible to
future experiments.
\par
On the other hand, various constraints imply that the scales
$M$ and $\VEV{S}$ cannot be arbitrarily low. At the scale
$M$, new colored multiplets appear which affect the running
of $\alpha_s$. Requiring that no Landau pole appears up
to some high energy scale gives a lower bound on $M$.
At the scale $\VEV{S}$ there are scalars which mediate
flavor changing neutral currents. Requiring that these
contributions would not exceed the experimental values
gives a lower bound on $\VEV{S}$. In this section we study
these bounds.
\par
The general formula for the running of $\alpha_s$
in our framework is
$$\(\alpha_s(M_P)\)^{-1}=\(\alpha_s(m_Z)\)^{-1}
+{7\over2\pi}\ln{M_{\rm SUSY}\over m_Z}
+{3\over2\pi}\ln{M\over M_{\rm SUSY}}
+{3-N_T\over2\pi}\ln{M_P\over M} \eqn\ourfram$$
where $M_{\rm SUSY}$ is the scale of the various sparticles and $N_T$ is
the number of pairs of extra color triplets and color anti-triplets.
The requirement that there is no Landau pole below the Planck mass $M_P$
can be translated into a lower bound on $M$ of the form
$$M_{\rm min}=M_P\exp(-Z/N_T), \eqn\mminmp$$
where $Z$ depends on $M_{\rm SUSY}$:
$$Z=\cases{174&$(i)\ M_{\rm SUSY}=m_Z$,\cr
183&$(ii)\ M_{\rm SUSY}=10m_Z$.\cr} \eqn\zmsusy$$
This leads to
$$\vbox{\settabs
\+ $N_f\quad$ & $\mu(\Lambda=30\ TeV)$ & $\mu(\Lambda=30\ TeV)$ \cr
\+ $N_T\quad$ & $M_{\rm min}^{(i)}\(TeV\)$
& $M_{\rm min}^{(ii)}\(TeV\)$ \cr
\+ $5$  & $7$           & $1.3$ \cr
\+ $6$  & $3\cdot 10^3$ & $6\cdot10^2$ \cr
\+ $7$  & $2\cdot 10^5$ & $4\cdot10^4$ \cr
\+ $8$  & $3\cdot 10^6$ & $1\cdot10^6$ \cr
\+ $9$  & $4\cdot 10^7$ & $1\cdot10^7$ \cr} \eqn\landsaf$$
We conclude that the new physics related to the scale $M$ may
be directly accessible to future experiments if $N_T\leq5$.
If $\VEV{S}\sim 0.04M$, then the new physics related to the
scale $\VEV{S}$ may be directly accessible if $N_T\leq5$
and could affect rare processes if  $N_T\leq7$.
Clearly, if the gauge group changes between $M$ and $M_P$ (e.g. $SU(3)$
is embedded in $SU(4)$) these bounds can be significantly weaker.

In our theoretical framework, there are necessarily scalars that
mediate FCNC at tree level. It is often thought that the
constraints on such scalars force them to be heavier than a
thousand $TeV$. This conclusion assumes that the relevant Yukawa
couplings of these scalars are of order one. Clearly, if the
couplings are smaller, the bounds are weaker. If the couplings of
these additional scalars are similar in their magnitude to those
of the Higgs which leads to masses, they must be small
$\(\wili-\hani\)$. In particular, it was pointed out in
ref. $\(\cheng\)$ that if the couplings of these additional
scalars are of the Fritzsch type they can be as light as a $TeV$.
A more complete analysis related to flavor symmetries was given
in ref. $\(\ahr\)$. There it is pointed out that the additional
scalars can be even lighter than a $TeV$.
\par
The most stringent bounds in our models arise from mixing of neutral
mesons. For all models discussed in the previous section, only
singlet fields $S_a$ contribute at tree level to meson mixing.
The effective nondiagonal Yukawa couplings of these fields are
typically of order
$$\lambda^a_{ij}\lambda^a_{ji}\sim{m_im_j\over\VEV{S_a}^2}.\eqn\scale$$
The mass of the field $S_a$ is also at the scale $\VEV{S_a}$.
Then the bounds of eq. \nfci\ can be translated into lower bounds
on the scale $\VEV{S}$ of the horizontal symmetry breaking:
$$\VEV{S}\gsim\cases{0.4\ TeV&($K-\bar K$ mixing),\cr
0.2\ TeV&($D-\bar D$ mixing),\cr
0.4\ TeV&($B-\bar B$ mixing).\cr}\eqn\scalef$$
Note that the bounds are very weak because the singlet mass
increases with $\VEV{S}$ while
its Yukawa couplings decrease with $\VEV{S}$.
\par
Additional bounds can be derived from rare decays such as
$B\rarrow X\mu^+\mu^-$, but they depend on details of the
leptonic sector, and are typically weaker than \scalef.
\par
We mentioned the possibility that the $U(1)_X$ symmetry
is broken, leading to non-renormalizable terms of the form \ntf.
In this case, the situation with FCNC is entirely different
as the scalar doublets contribute as well. The nondiagonal Yukawa
couplings of the doublets in this class of models are typically
of order
$$\lambda^q_{ij}\lambda^q_{ji}\sim{m_im_j\over\VEV{\phi_q}^2}.
\eqn\scali$$
Taking $\VEV{\phi_q}\sim0.2\ TeV$, we find from eq. \nfci:
$$M(\phi_q)\gsim\cases{0.8\ TeV&($K-\bar K$ mixing),\cr
0.3\ TeV&($D-\bar D$ mixing),\cr
0.7\ TeV&($B-\bar B$ mixing).\cr}\eqn\scalj$$
The strongest bound, coming from $K-\bar K$ mixing, is in conflict
with the unitarity bound on the Higgs mass of 0.75 $TeV$ (for a review
see ref. $\(\ghkd\)$). Strictly
speaking, this bound applies to the single Higgs of the minimal
standard model, but more generally it applies approximately to the
doublet scalar that carries the VEV. It is even more difficult to
accommodate \scalj\ in a supersymmetric framework. There the upper
bound on the mass of the lightest neutral scalar is of order
0.15 $TeV$. Again, strictly speaking the bound applies only
in the minimal supersymmetric standard model, but if we require
that perturbativity holds to high energy scales, this bound becomes
very general $\(\hash,\kkw\)$.
\par
A second problem in models with no $U(1)_X$ symmetry arises if
some of the quark sector parameters depend on the ratio
defined in eq. \ntg, $\eta^2={\VEV{\phi_d}\VEV{\phi_u}\over
M^2}$. The experimental value of the parameters will then set
the scale $M$. For example, if $\eta^2$ takes the role of
our small parameter $\e_2$, namely $\t^2\sim0.008$, then
$M\lsim2\ TeV$ which may be in conflict with the bounds from
Landau poles, or force the scale $\VEV{S}$ to be too low
for FCNC bounds. Furthermore, the light fermions would have exotic
components of order ${\VEV{\phi}\over M}\sim0.1$ which is inconsistent
with various electroweak precision measurements $\(\nrt\)$.
If $\t^2$ just explains the first generation masses,
namely $\t^2\sim{m_d\over m_b}$ or ${m_u\over m_t}$, then it
could be useful in explaining these parameters without forcing
$M$ to be too small.
\par
A third problem arises in models of broken $U(1)_X$
where the hierarchy $m_b\ll m_t$ results from $\VEV{\phi_d}\ll
\VEV{\phi_u}$. If, for example, $\VEV{\phi_d}\sim0.1\VEV{\phi_u}$,
then the bounds on $M(\phi_d)$ are about ten times stronger
than those in \scalj. This is not necessarily in conflict with
the stringent supersymmetric upper bound mentioned above, because
in this case the lightest neutral scalar is dominantly $\phi_u$.
But in supersymmetric models it is in conflict with perturbativity,
while in non-supersymmetric models, it sets a lower bound on the
scale $\VEV{S}$ beyond the direct reach of future experiments.
\par
We should mention that even in models which are $U(1)_X$-symmetric,
the light scalars may mediate FCNC because they have small
components of the singlet fields. However, as these components
are of order $\VEV{\phi}^2/\VEV{S}^2$, the contributions to neutral
meson mixing are
usually smaller than those from the heavy scalars which are
dominantly singlets $\(\hani\)$.
\par
To summarize, models with $U(1)_X$ symmetry and a small number
of massive colored supermultiplets allow for rich phenomenology
at energies accessible to future experiments - a $TeV$ or even lower.
Models with $U(1)_X$ symmetry and a large number of massive
colored supermultiplets can explain all the details of the
observed hierarchy in the quark sector parameters, but have no
directly accessible phenomenology (similarly to the original
Froggatt-Nielsen models $\(\frni\)$). Supersymmetric
models with no $U(1)_X$ symmetry are probably not viable.
\chapter{Conclusions}
The new data on $m_t$, $V_{cb}$ and $V_{ub}$ motivated us to
reexamine the old problem of the quark mass matrix.  Perhaps the most
puzzling aspect of the quark mass matrix is the large
hierarchy between the entries.  Therefore, as a first
approximation we are not interested in explaining the precise values of
the parameters in the mass matrix but focus on the order of magnitudes.

We parametrize the mass matrices as
$$M_u=\VEV{\phi}\pmatrix{\e_2^2&0&\e_2\cr
\e_1\e_2&\e_2&\e_1\cr \e_2&0 &1\cr},
\qquad\qquad
M_d=\VEV{\phi}\pmatrix{\e_2^2&\e_1\e_2&\e_1\e_2\cr
\e_1\e_2&\e_1^2&\e_1^2\cr \e_2&\e_1&\e_1\cr}.\eqn\explcc$$
leading to
$$\eqalign{
\e_1&\sim|V_{cb}|,\ {m_s\over m_b},\  {m_b\over m_t}\cr
\e_2&\sim|V_{ub}|,\ {m_c\over m_t},\  {m_u\over m_c} \cr
{\e_2\over\e_1}&\sim|V_{us}|,\ \sqrt{m_d\over m_s}\cr
}\eqn\finres$$
Experimentally, these relations are satisfied quite well with $\e_1 \sim
0.04 $ and $\e_2 \sim 0.008$.  These are small but not extremely small
numbers.  The large hierarchy occurs by raising these numbers to large
powers.

Following [\frni] and [\dimo] we attempt to explain the hierarchy in
\explcc\ in terms of tree level exchanges of massive particles with
mass of order $M$ and the small parameters $\e_1={\VEV{S_1} \over M}$
and $\e_2={\VEV{S_2} \over M}$ are related to the expectation values of
two massive scalars (we absorb a Yukawa coupling of $S$ to massive
fields into its VEV).  This is most easily achieved if we assume that
$m_u=0$ and there are two Higgs fields with $\VEV{\phi_d}\sim \e_1
\VEV{\phi_u}$.  Then the matrices in \explcc\ can be replaced with
$$M_u=\VEV{\phi_u}\pmatrix{0&0&0\cr
0&\e_2&\e_1\cr 0&0 &1\cr},
\qquad\qquad
M_d=\VEV{\phi_d}\pmatrix{0&\e_2&\e_2\cr
\e_2&\e_1&\e_1\cr 0&1&1\cr}.\eqn\explccs$$
A high energy theory containing a single $U$-like quark and a couple of
$D$-like quarks can lead to \explccs.

A more detailed description with a non-zero $m_u$ and an explanation of
the hierarchy between $\VEV{\phi_u}$ and $\VEV{\phi_d}$ needs two more
$U$-like quarks and three more $D$-like quarks in the high energy theory.
Such a theory reproduces all the relations in \finres.

The parametrization \explcc\ is not unique.  Other parametrizations are
possible and they lead one to consider different high energy theories.
Our models should therefore be viewed merely as an existence proof to
the approach presented here.  In constructing a complete model one needs
to find the discrete horizontal symmetry (rather than pretend
that it is continuous) and make sure that it does not suffer from
anomalies.

Previous attempts to explain the hierarchy in the quark mass matrices in
terms of tree level exchanges of massive fields [\frni, \dimo] assumed
that the relevant flavor dynamics occurs at super high energies.  Both
$\VEV{S_i}$ and $M$ are very large ($10^{10} -10^{19} GeV$) with a small
(but not extremely small) ratio between them $\e_i$.  Unlike these
authors, we suggest that the entire flavor dynamics can take place at
experimentally accessible energies  in the TeV range.

It is amusing to note that the set of particles that we need to add to
the standard model are common in string inspired models.  There we
typically have a number of generations and anti-generations containing
pairs of mirror fields with the quantum numbers of the standard quarks
and leptons.  Furthermore, the $E_6$ generations contain also pairs of
mirror singlet $D$ quarks and pairs of mirror lepton doublets as well as
fields like our $S$.  Also,
string models often have large discrete symmetries which could be used
as horizontal symmetries.  It is fortunate that unlike grand unified
theories these discrete symmetries can act differently on different
fields in the same generation (in fact it is not clear how to group the
fields into generations).  The particle content of a string inspired
model with four generations and a single anti-generation is large enough
to produce all the hierarchies we need except that  $m_u=0$.  Although
with more generations it is possible to induce a non-zero mass for the
up quark we would like to remind the reader that $m_u=0$ leads to a
possible solution to the strong CP problem and may also be consistent
[\kama-\choi].

\bigskip
\centerline{\bf Acknowledgements}

It is a pleasure to thank T. Banks, A. Dabholkar, M. Dine, M.C.
Gonzalez-Garcia and A. Schwimmer for several useful discussions.  This
work was supported in part by DOE grant DE-FG05-90ER40559.  Y.N. is an
incumbent of the Ruth E. Recu Career Development Chair, supported in
part by the Israel Commission for Basic Research, and by the Minerva
Foundation.

\bigskip
\centerline{\bf Appendix: Lepton Masses}

The observed hierarchy in the charged lepton masses can be explained by
a similar mechanism to the quark sector parameters.  The experimental
values of the masses are
$$m_e=0.51\ MeV,\ \ m_\mu=105.7\ MeV,\ \ m_\tau=1777\ MeV,\eqno{(A.1)}$$
leading to
$${m_e\over m_\mu}=0.005,\ \
{m_\mu\over m_\tau}=0.06,\ \ {m_\tau\over
m_t}=0.008^{+0.004}_{-0.002}.\eqno{(A.2)}$$
(In this work we take the neutrinos to be massless.)
\par
With the two small parameters defined in section 6,
$\e_1\sim0.04$ and $\e_2\sim0.008$, we should aim at
$$\eqalign{\e_1\sim&{m_\mu\over m_\tau},\cr
\e_2\sim&{m_e\over m_\mu},\ {m_\tau\over m_t}.\cr}\eqno{(A.3)}$$
We should point out, however, that other possibilities exist.
With this order of magnitude estimate, we have
$$\det M_\ell\sim\VEV{\phi_d}^3\e_1^2\e_2^4,\eqno{(A.4)}$$
requiring at least six heavy charged leptons.
\par
We denote leptons doublets by $L_i$ and charged lepton singlets
by $\ell^+_i$ and $E^\pm_i$ for light and heavy fields, respectively.
We use the same set of scalar fields as in eq. \explca.
To produce the mass ratios (A.3), we take
$$\vbox{\settabs\+ {\rm field} $\quad$ &
$\quad- {2 \over 3}$  & $\quad x -2$ \cr
\+ \rm {field} & $ H_1\quad$  & $H_2\quad $ \cr
\+ $L_1$ & $1$  & $1$ \cr
\+ $L_2$ & $1$  & $0$ \cr
\+ $L_3$ & $0$  & $0$ \cr
\+ $\ell^+_{1,2,3}$  & $0$ & $1$ \cr
\+ $E^-_{1,2,3}   $  & $0$ & $0$ \cr
\+ $E^-_{4,5}     $  & $1$ & $0$ \cr
\+ $E^-_6         $  & $1$ & $1$ \cr
}\eqno{(A.5)}$$
and $E^+_i$ fields with conjugate horizontal and gauge numbers
to those of $E^-_i$.
\par
In the full theory we have a $9\times9$ mass matrix whose rows
correspond to $\{L_1,L_2,L_3,E^-_1,E^-_2,E^-_3,E^-_4,E^-_5,E^-_6\}$
and its columns to
$\{\ell^+_1,\ell^+_2,\ell^+_3,E^+_1,$ $E^+_2,$
$E^+_3,E^+_4,E^+_5,E^+_6\}$.
Using the quantum numbers in (A.5) we find the orders
of magnitude for the various entries:
$$M_\ell^{\rm full}=\pmatrix{
0&0&0&0&0&0&0&0&\VEV{\phi_d}\cr
0&0&0&0&0&0&\VEV{\phi_d}&\VEV{\phi_d}&0\cr
0&0&0&\VEV{\phi_d}&\VEV{\phi_d}&\VEV{\phi_d}&0&0&0\cr
\VEV{S_2}&\VEV{S_2}&\VEV{S_2}&M&0&0&0&0&0\cr
\VEV{S_2}&\VEV{S_2}&\VEV{S_2}&0&M&0&0&0&0\cr
\VEV{S_2}&\VEV{S_2}&\VEV{S_2}&0&0&M&0&0&0\cr
0&0&0&\VEV{S_1}&\VEV{S_1}&\VEV{S_1}&M&0&0\cr
0&0&0&\VEV{S_1}&\VEV{S_1}&\VEV{S_1}&0&M&0\cr
0&0&0&0&0&0&\VEV{S_2}&\VEV{S_2}&M\cr}.\eqno{(A.6)}$$
After integrating out the heavy fields, we get for the three light
generations
$$M_\ell=\VEV{\phi_d}\pmatrix{\e_2^2\e_1&\e_2^2\e_1&\e_2^2\e_1\cr
\e_2\e_1&\e_2\e_1&\e_2\e_1\cr \e_2&\e_2&\e_2\cr},\eqno{(A.7)}$$
leading to the mass ratios of (A.3).  Clearly, some of the six
massive $E$'s can be replaced with massive $SU(2)$ doublets with
appropriate horizontal charges without affecting the low energy
hierarchy.
\par
If we explain the small ratio $m_\tau/m_t$ dynamically, namely
$m_\tau/m_t \sim \VEV{\phi_d}/\VEV{\phi_u}$, then
$\det M_\ell\sim\VEV{\phi_d}^3\e_1^2\e_2$. This can be derived
from a model with three heavy charged leptons only.
An example of such a model is obtained by modifying the $H_2$ charges
of the $\ell^+_i$ fields of the model above to
$$H_2(\ell^+_i)=0,\eqno{(A.8)}$$
and by dropping the massive leptons $E_1$, $E_2$ and $E_3$.
After integrating out the heavy fields we find
$$M_\ell=\VEV{\phi_d}\pmatrix{\e_2\e_1&\e_2\e_1&\e_2\e_1\cr
\e_1&\e_1&\e_1\cr 1&1&1\cr}.\eqno{(A.9)}$$
\par
Finally, let us mention that the bounds on the masses of
scalars with non-diagonal couplings to charged lepton mass
eigenstates are
$$M(\phi_i)\gsim\cases{
100\ TeV\ \sqrt{\lambda^i_{e\mu}\lambda^i_{ee}}&($\mu\rarrow eee$),\cr
1\ TeV\ \sqrt{\lambda^i_{\tau\ell}\lambda^i_{\ell\ell}}
&($\tau\rarrow\ell\ell\ell$).\cr}\eqno{(A.10)}$$

\refout
\end